\documentclass[preprint,12pt]{elsarticle}

\usepackage{amsmath}
\usepackage{amssymb}
\usepackage{graphicx}
\usepackage{multirow}
\usepackage{multicol}

\DeclareGraphicsExtensions{.png,.pdf,.jpg} 

\usepackage{textcomp} 

\usepackage{lineno}

\usepackage[
pdfauthor={Olivier Bourrion},
pdftitle={C2D8: An eight channel CCD readout electronics dedicated to low energy neutron detection.},
pdfsubject={C2D8},
pdfkeywords={},
colorlinks={true},
linkcolor=red,
citecolor=green,
urlcolor=cyan,
pdfstartview={FitH}
]{hyperref}

\begin{document}
\begin{frontmatter}
 
\title{C2D8: An eight channel CCD readout electronics dedicated to low energy neutron detection.}
\author[LPSC]{O.~Bourrion\corref{cor1}}
\ead{olivier.bourrion@lpsc.in2p3.fr}
\author[LPSC]{B.~Clement}
\author[LPSC]{D.~Tourres}
\author[LPSC]{G.~Pignol}
\author[LPSC]{Y.~Xi}
\author[LPSC]{D.~Rebreyend}
\author[ILL]{V.V.~Nesvizhevsky}

\cortext[cor1]{Corresponding author}
\address[LPSC]{LPSC, Universit\'e Grenoble-Alpes, CNRS/IN2P3 \\
53, avenue des Martyrs, Grenoble, France}
\address[ILL]{ILL, Institut Laue Langevin\\
71, avenue des Martyrs, Grenoble, France}

\begin{abstract}
Position-sensitive detectors for cold and ultra-cold neutrons (UCN) are in use in fundamental research. 
In particular, measuring the properties of the quantum states of bouncing neutrons requires micro-metric spatial resolution. 
To this end, a Charge Coupled Device (CCD) coated with a thin conversion layer that allows a real time detection of neutron hits is under development at LPSC.
In this paper, we present the design and performance of a dedicated electronic board designed to read-out eight CCDs simultaneously and operating under vacuum.
\end{abstract}


\end{frontmatter}
\section{Introduction}
\label{introSec}
Neutrons interact with matter mostly through strong nuclear interaction. When the neutron wavelength becomes commensurate with the inter-atomic spacing, only coherent scattering occurs.
As a consequence, neutrons slowed down to kinetic energies below 100\,neV are totally reflected at any angle of incidence by most solid surfaces. These \emph{ultra-cold neutrons} constitute a sensitive tool to study fundamental interactions and symmetries \cite{Dubbers2011}. 
In particular, ultra-cold neutrons are used to study gravity in a quantum context \cite{Nesvizhevsky2002,Nesvizhevsky2010,Jenke2011,Pignol2015}.  
A neutron bouncing on top of an horizontal mirror realizes a simple one dimensional quantum well problem and the vertical motion of the neutron bouncer has discrete energy states. 
The wave functions associated to the stationary quantum states have a spatial extension governed by the parameter $z_0 = (\hbar^2/2m_n^2g)^{1/3} \approx 6$\,\textmu m. 
Therefore, observing the spatial structure of the quantum states requires position-sensitive neutron detectors with a micro-metric spatial resolution. 

Semiconductor-based detectors, coated with adequate neutron converters, have been demonstrated to be well suited for this kind of measurements \cite{Jakubek2009,Jakubek2009b,Kawasaki2010,Lauer2011}.
The UCNBox (Ultra Cold Neutrons BOron piXels) detector has been recently developed as position sensitive sensor optimized to measure the wave-functions of the bouncing neutron in the GRANIT experiment \cite{Roulier2015}.
The GRANIT facility uses a 30\,cm wide glass mirror as the surface where neutrons bounce.
The setup is inside a vacuum chamber at $10^{-5}$\,mbar.
The detector, composed of 8 Charge Coupled Devices (CCD), is designed to cover a sensitive area of $300 \, {\rm mm} \times 0.8 \, {\rm mm}$.
Each CCD is an Hamamatsu S11071-1106N sensor (pixel size 14\,\textmu m~$\times$~14\,\textmu m and number of effective pixels $2048 \times 64$) coated with $^{10}$B, thanks to plasma assisted physical vapor deposition~\cite{boronCCD}.

Neutron capture on boron produces in most cases both a 1.5\,MeV $\alpha$ particle and a 0.8\,MeV $^{7}$Li nucleus.
The CCD sensor is used as a pixelated silicon detector.
The charge produced by the energy deposition migrates in the neighboring pixels allowing a precise reconstruction of the position using weighted average.
To limit this migration of charges each CCD sensor must be read at approximately a 1\,Hz rate, with a dead-time as low as possible.
In this project we aimed at dead-times below 1\,\%.
The UCN rate in the final experiment should not exceed 1\,Hz per sensor. Nevertheless, for calibration purposes, rates as high as 50\,Hz per sensor are desirable.
Also, given the fact that no mechanical shutter system can be used to avoid neutron detection, the shortest possible readout time must be achieved for each CCD to avoid neutron detection during the CCD charge transfer, as it would corrupt the data.
Consequently, the CCD must be read-out at the maximum speed specified by the manufacturer (10\,MHz), they can however be read one after another.

Typically, both detection-rate and number of hit pixels are low as all charges from one charged particle are collected within a $11\times11$ pixels matrix. For a maximum rate of 50\,Hz per sensor, only 5\,\% of the sensor contains useful data.
In normal conditions, this drops to 0.1\,\%.
This calls for the implementation of a data reduction system, that removes any pixel data below a discrimination threshold.

It must be noted, that an adjustable exposure time must be implemented to permit the use of bright light sources such as LED used to test and adjust CCD alignment.

Additionally, the readout system must be located very close to the sensor system in order to minimize signal integrity issues, but also to minimize the number of vacuum feed-through for the CCD signals (control and readout).
This requirement, implies that the readout must be located inside the detector cell and thus withstand vacuum conditions.
Consequently, special care must be taken on power usage and heat dissipation to ensure proper operation.
This paper is organized as follows: section~\ref{HardwareSec} presents the hardware design, section~\ref{FPGASec} describes the firmware architecture.
Eventually, a short summary is given in section~\ref{SummarySec}.

\section{Hardware description}
\label{HardwareSec}
\begin{figure}[ht]
\begin{center}
\includegraphics[angle=0,width=0.9\textwidth]{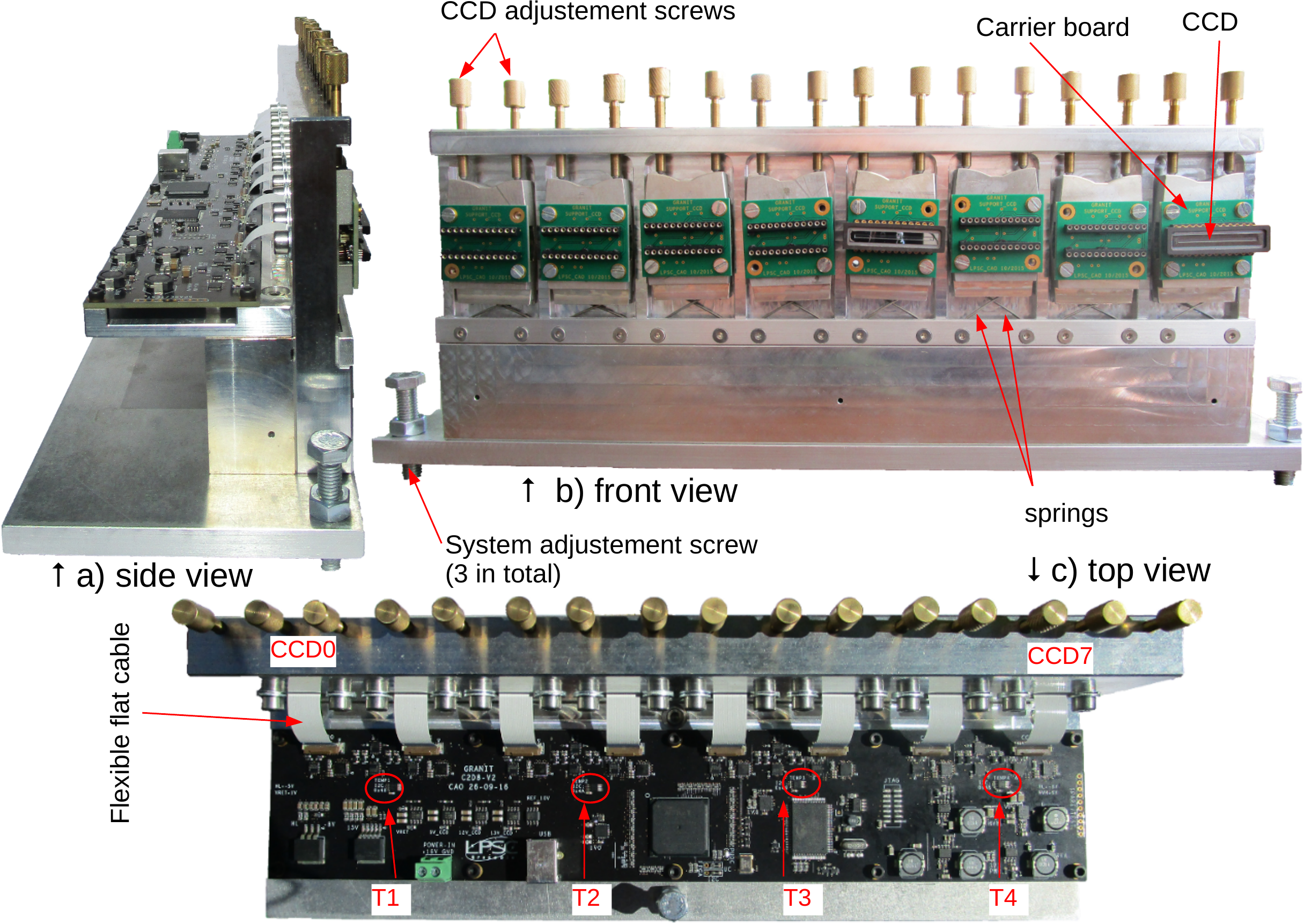}
\caption{Picture of the readout electronics mounted on its mechanical support.
The electronic system is composed of two parts: the front-end composed of 8 carrier boards, each holding a CCD, and the back-end with the control and readout circuits. 
The boards are mounted on a mechanical support system equipped with adjustment screws to allow the adjustment of each CCD carrier board position.
The four thermal sensors implemented on the board are indicated (T1 to T4).
}
\label{c2d8Pic}
\end{center}
\end{figure}

To meet the requirements listed in section~\ref{introSec}, we have opted for a solution based on two distinct modules: a front-end part composed of 8 boards, each holding a CCD; a back-end part for the control and readout circuits.
As shown in figure~\ref{c2d8Pic}, the front-end (FEB) and back-end boards are mounted on a common mechanical support.
This support is designed such as to be able to finely adjust the position of each CCD board, thanks to dedicated screws.
We have checked that this system allows for a relative alignment of the 8 CCDs within 10\,\textmu m, sufficient for our needs.

Additionally, the electronics and the support system were designed to optimize the thermal coupling.
Indeed, the support system is used to conduct a significant part of the heat flow to the vacuum vessel, while the remaining part of the heat is radiated in the chamber.

\begin{figure}[ht]
\begin{center}
\includegraphics[angle=0,width=0.9\textwidth]{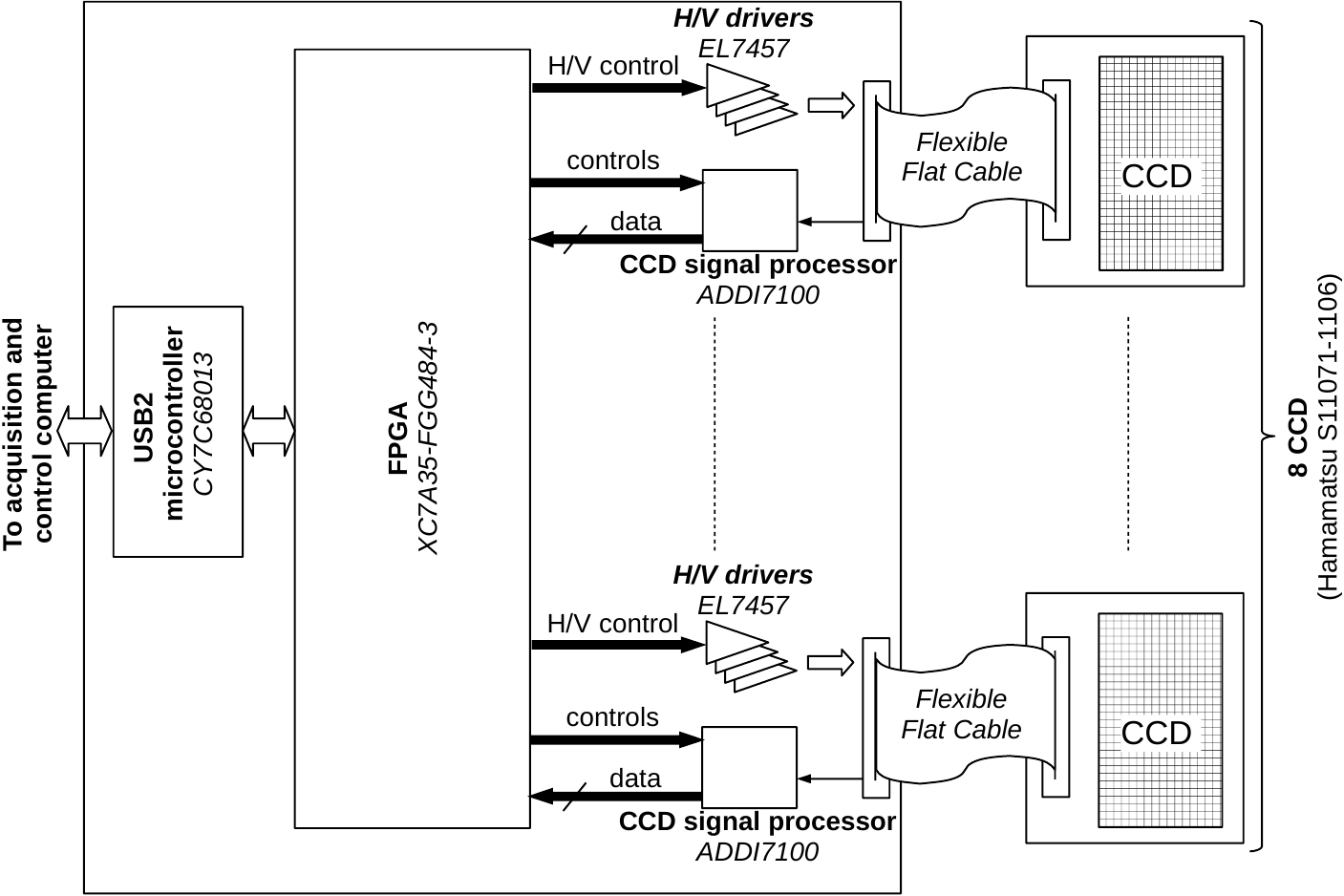}
\caption{Block diagram of the electronics system. 
Each FEB is connected to the back-end board with a Flexible Flat Cable (FFC). 
The back-end board is in charge of generating the CCD control signals (horizontal/ vertical shifts and reset gate), to perform the CCD signal digitization, to aggregate the data and to finally make them available for readout via a Universal Serial Bus (USB) interface.
Each CCD signal digitization is done by a dedicated CCD signal processor \cite{ADDI7100}.}
\label{elecDiag}
\end{center}
\end{figure}

A block diagram of the electronics is shown in figure~\ref{elecDiag}.
Each FEB is connected to the back-end board with a Flexible Flat Cable (FFC). 
The back-end board is in charge of generating the adequate CCD control signals (horizontal/ vertical shifts and reset gate); performing the CCD signal digitization; aggregating the data and finally making them available for readout via a Universal Serial Bus (USB) interface.

Each CCD signal digitization is carried-out by a dedicated CCD signal processor (Analog devices ADDI7100  \cite{ADDI7100}).
This CCD processor can operate at 45\,MHz, which is significantly faster than the maximum readout speed of the Hamamatsu S11071-1106 CCD (10\,MHz), and has a digitization resolution of 12 bit with noise performance better than the CCD performance.

Indeed, the CCD processor is specified for having a system noise equivalent to 24.4\,e\textsuperscript{-} (0.8\,LSB~rms with CDS gain set at +6\,dB for a typical CCD sensitivity of 8\,\textmu V/e- which corresponds to 30.5e\textsuperscript{-}/ADU) while the typical CCD noise is composed of the readout noise (23\,e\textsuperscript{-}~rms) and of the dark current integration (typically 50\,e\textsuperscript{-}/pixel/s at 25\textdegree C) resulting in a total of 73\,e\textsuperscript{-} for 1\,s of integration.
The total expected system noise is thus dominated by the CCD and is about 77\,e\textsuperscript{-}, which is compatible with the measurements that showed a system noise of 91.5$\pm$15\,e\textsuperscript{-}.

The control signals required to read-out the CCD are generated by a Field Programmable Gate Array (FPGA).
These signals, composed of Horizontal/Vertical (H/V) shifting signals and Reset Gate signals (RG), are amplified by dedicated drivers to accommodate the CCD load.
The FPGA is also used to produce the signals necessary to operate the CCD signal processor, i.e. clamping and pre-blank signals, the correlated double sampler  (CDS) signals and the serial control links.

The rationales for selecting the FPGA used in this design (Xilinx XC7A35-FGG484) were (i) its low power consumption; (ii) the possibility to precisely adjust the timing of the generated signals; (iii) the large amount of memory available.
Indeed, the power had to be minimized by design as much as possible to permit the electronics operation under vacuum while avoiding a too fancy mechanical setup for thermalization (for instance: usage of the standby modes of the CCD processors during the exposure time).
Additionally, this FPGA features high performance serializers in each of its input/output block.
Thanks to these blocks, one can adjust output signals with a time resolution of about 2\,ns by using a high speed 480\,MHz clock (see section~\ref{FPGASec}).
This makes it possible to conveniently set the CDS sampling times.
The sizing of the memory was based on the criteria that its capacity should be at least half of the memory required to buffer the data generated by one CCD readout, i.e. $2048 \times 64 /2=65536$ 16-bits words, corresponding to more than 1\,Mbit of storage.
This time equivalent buffering must be considered acknowledging the fact that, by specification, a new USB2 transaction can be placed every millisecond.

\begin{figure}[ht]
\begin{center}
\includegraphics[angle=0,width=0.8\textwidth]{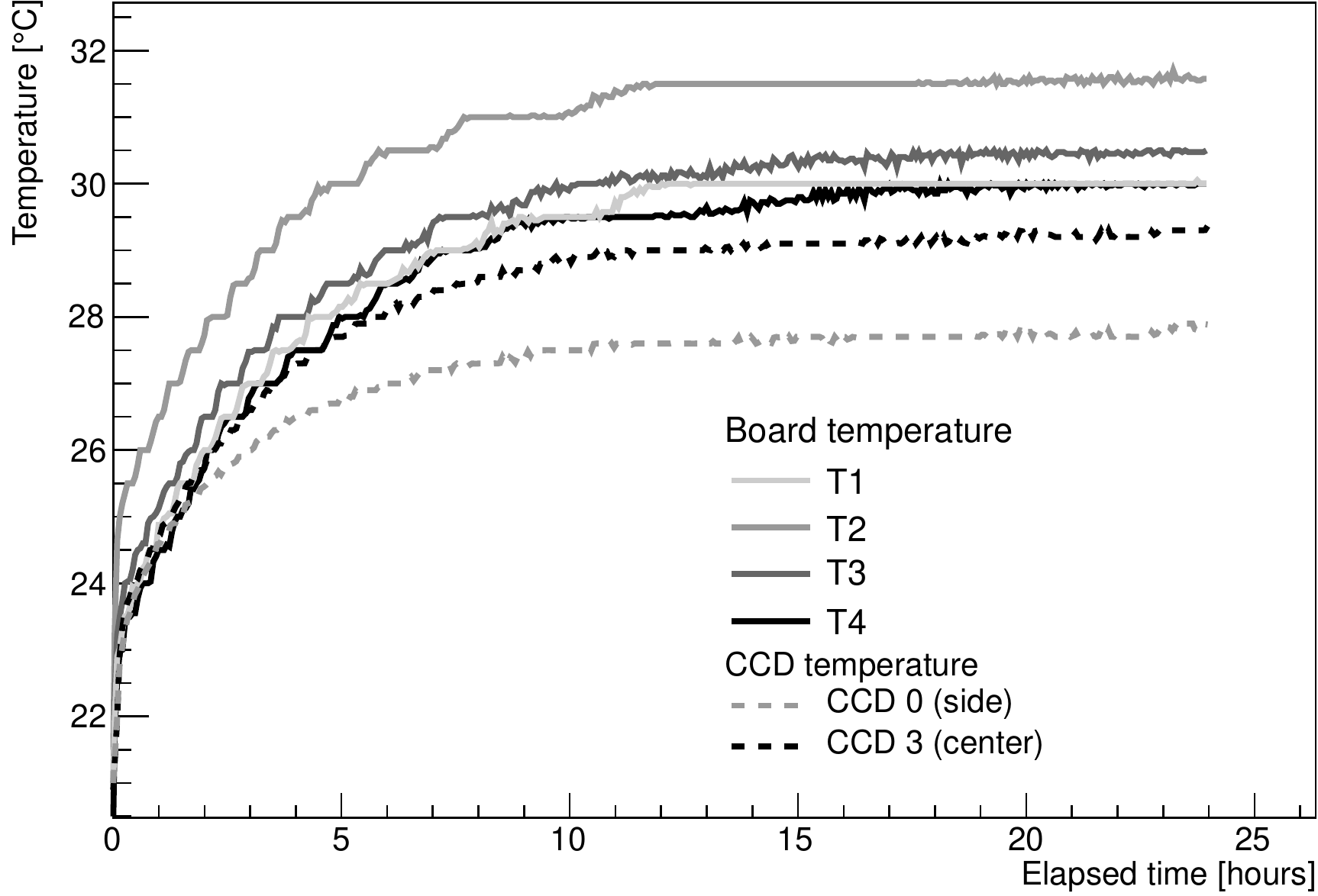}
\caption{Plot of the system temperatures recorded from power-up to equilibrium.
We can see that the CCD highest temperature elevation from ambient is about 8\textdegree C.
The highest temperature elevation of 11\textdegree C is measured for probe T2 which is located close to the FPGA.
}
\label{ccdTemp}
\end{center}
\end{figure}

The total system power usage was measured to be 3\,W in full readout.
In this budget, we estimate by combining specifications and measurements that about 140\,mW are used for each CCD (including about 32\,mW for the biasing and the line drivers losses); 600\,mW are used by the power converters (linear and switching) distributed on the board; about 1200\,mW are used by the FPGA.
To asses the operating temperature of the system, dedicated measurements were performed under vacuum.
A total of six temperatures were recorded over 25 hours, i.e. from the system power-up until system equilibrium (see figure~\ref{ccdTemp}).
The temperature of the back-end electronics was recorded on four points thanks to the sensor circuits included in design (LM75B), see locations in figure~\ref{elecDiag}
Additionally, the temperature of two CCD were recorded: one at the border and one in the middle of the detector plane with thermally coupled PT100 probes.
We can see that the CCD highest temperature elevation from ambient is about 8\textdegree C.
The highest temperature elevation of 11\textdegree C is measured for probe T2 which is located close to the FPGA.

\section{Firmware description}
\label{FPGASec}
\begin{figure}[ht]
\begin{center}
\includegraphics[angle=0,width=0.9\textwidth]{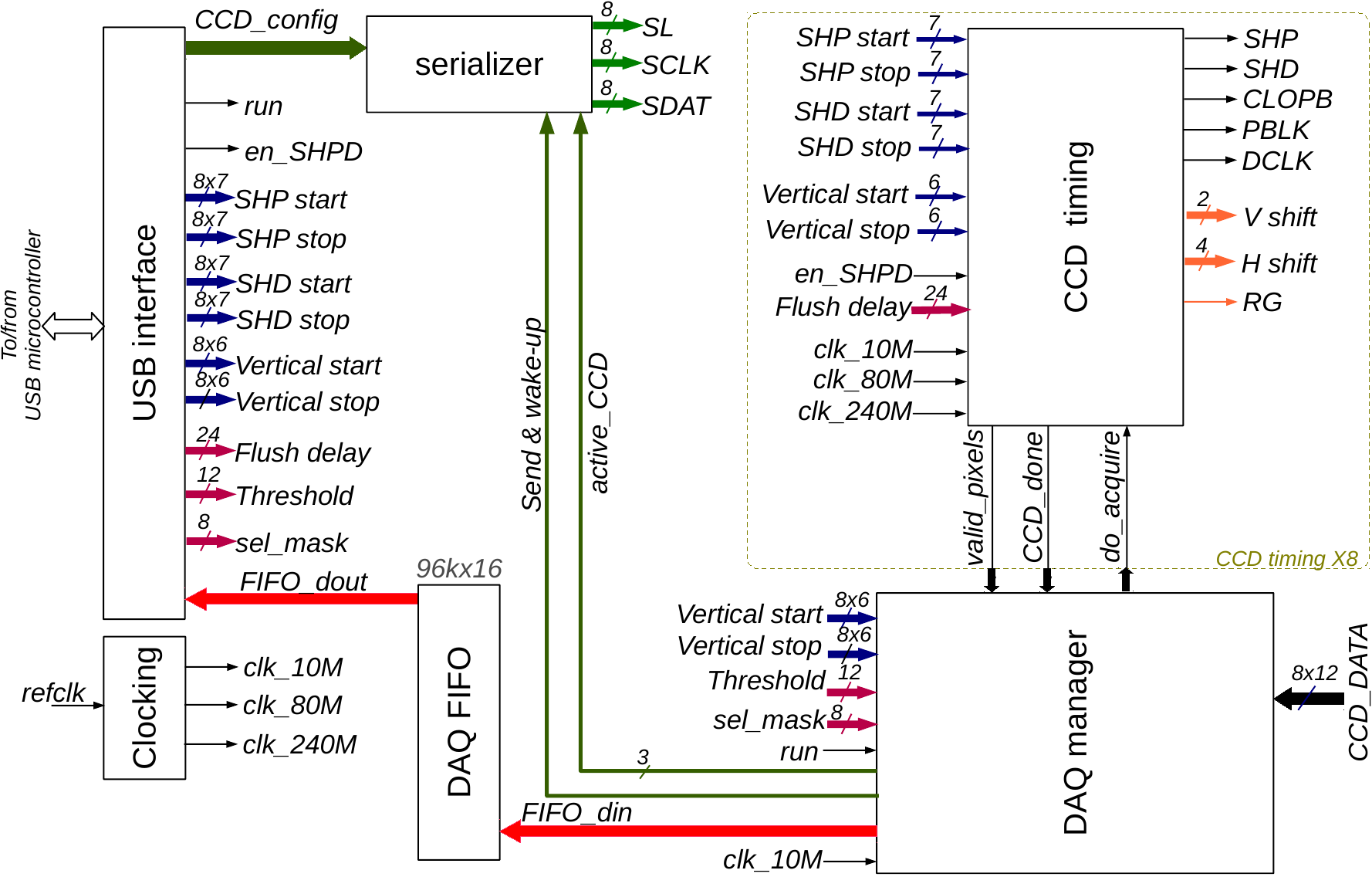}
\caption{Block diagram of the FPGA firmware. It is composed of six different blocks: clocking, USB interface, serializer, DAQ FIFO, DAQ manager and CCD timing.
}
\label{firmDiag}
\end{center}
\end{figure}

A block diagram of the FPGA firmware is shown in figure~\ref{firmDiag}.
It is composed of six different blocks: `clocking', `USB interface', `serializer', `DAQ FIFO', `DAQ manager' and `CCD timing'.
They are described hereafter.

The `clocking' module is used to produce the CCD readout clock (10\,MHz) and the fast clocks required by the timing modules (respectively 80\,MHz and 240\,MHz). 
To achieve this, a Mixed Mode Clock Manager \cite{MMCM} (MMCM) uses the 50\,MHz reference clock, provided at the board level from a crystal oscillator, to perform the clock generation.

The `USB interface' provides an interface between the USB micro-controller and the various configuration registers as well as an interface to read-out the acquisition FIFO (`DAQ FIFO'). 

The `DAQ FIFO' is designed as a First Word Fall Through FIFO, it provides a buffering depth of 98305 words of 16-bit.
This configuration was chosen to fully exploit the available memory in the selected FPGA and thus loosen the constraints on the acquisition software.

The `serializer' module is used to interface the CCD processor serial configuration link, it is controlled either by the `USB interface' or by the `DAQ manager'.
When controlled by the `USB interface', it allows the acquisition and control software to configure the CCD processor with the required parameters (gain settings, operation mode, ...).
During acquisition mode, the `DAQ manager' can use the `serializer' to change the CCD processor operating mode (normal operation or full standby) of the active CCD.
Hence, the CCD processors are in normal mode only when required, less than 1\% of the time, and thus the power used is minimized.

The eight `CCD timing' modules, one per CCD, are used to generate the CCD shifting signals (vertical, horizontal and reset gate) and the CCD processor control signals.
These are the pedestal and data sampling control signals required for the correlated double sampler (respectively SHP and SHD), the preblank signal (PBLK) which is used to clear the processor output data during vertical shifting and the clamp optical black signal (CLOPB) used to remove residual offsets in the CCD processor signal chain.
The CLOPB signal is supposed to be activated when the black pixels are being read-out.
A detailed description of the `CCD timing' module is given in section~\ref{ccdTimingSec}.

The `DAQ manager' module performs two main tasks.
Its first role is to sequence the CCD acquisition by triggering the appropriate `CCD timing' module.
The CCD to read-out are individually selected by an eight bit selection mask (`sel\_mask').
Its second purpose is to recover the data provided by the CCD processor associated with the CCD being accessed, to discriminate the data with respect to a threshold, to encapsulate the data and to eventually store them in the output buffer.
More details about the `DAQ manager' is given in section~\ref{daqManagerSec}.

\subsection{CCD timing module description}
\label{ccdTimingSec}
\begin{figure}[ht]
\begin{center}
\includegraphics[angle=0,width=0.9\textwidth]{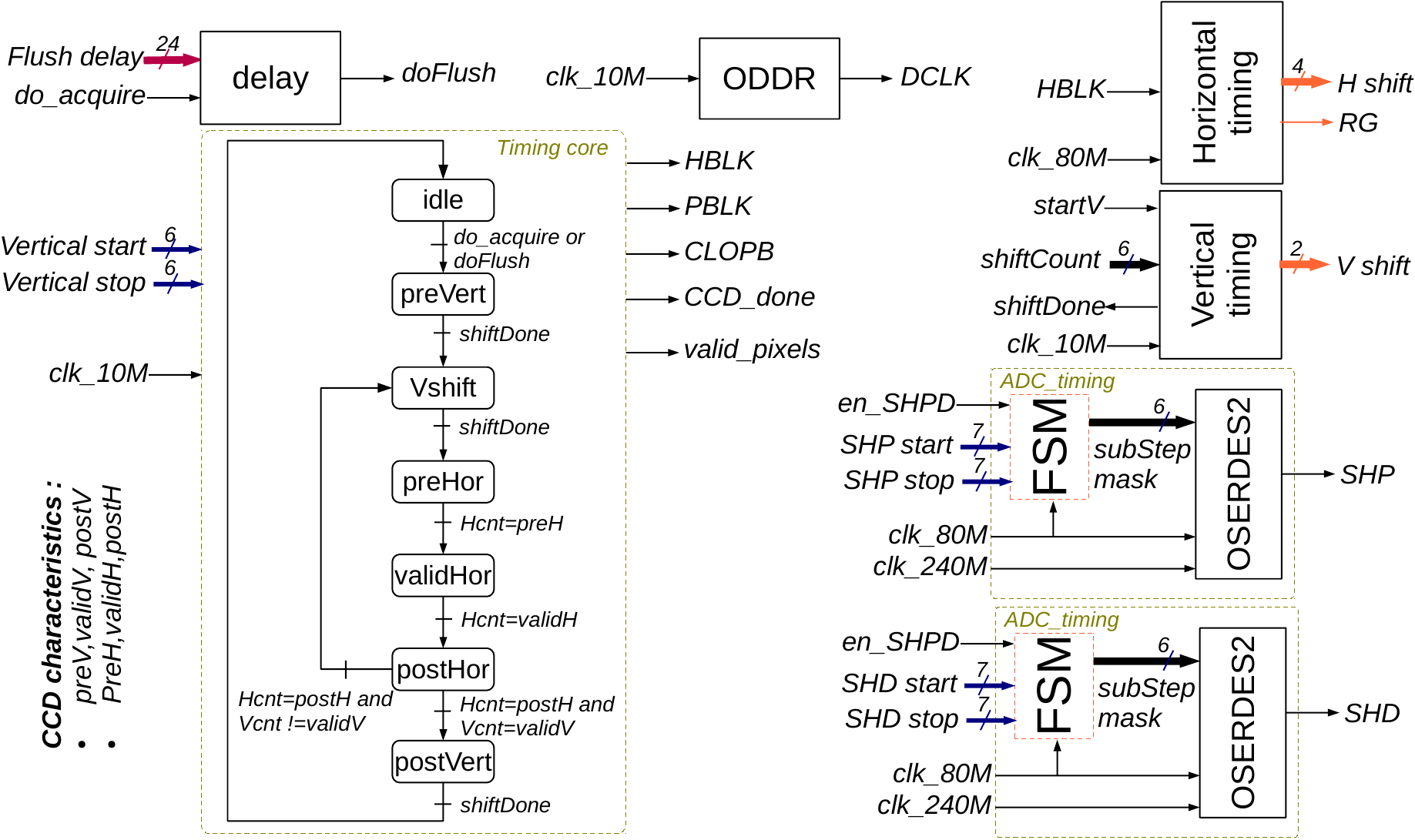}
\caption{Block diagram of the CCD timing module.}
\label{timingBlockDiag}
\end{center}
\end{figure}

A block diagram detailing the internal architecture of the `CCD timing' module is given in figure~\ref{timingBlockDiag}.
The `horizontal timing' module is in charge of controlling the four horizontal shifts (H shift) and the reset gate (RG) signals.
Module operation is deactivated by the Horizontal Blank signal (HBLK), that is when the `timing core' FSM is not in the horizontal shifting states (\textit{preHor}, \textit{validHor} and \textit{postHor}).
The horizontal shifting are done at the pixel clock speed, i.e. 10\,MHz. 
To cope with the various signal phases and widths required by the chosen CCD, the module is clocked eight times faster than the pixel clock.

The `vertical timing' module controls the two vertical shifting signals.
The module is designed to be able to shift several lines before signaling completion with \textit{shiftDone}.
The vertical shifting is requested by \textit{startV} and the number of lines to move is determined by \textit{shiftCount}.
It may be noted that the CCD is operated in the Large Saturation Charge Mode  \cite{LSCM}.

The `timing core' Finite State Machine (FSM) is used to coordinate the CCD control signals.
It controls the horizontal and vertical timing modules and thus activates them when appropriate.
The FSM is started either by the acquisition signal (\textit{do\_acquire}) or by the flush request signal (\textit{doFlush}), which is a delayed version of the \textit{do\_acquire} signal.
Note that by design, no data is recorded when the FSM is triggered by the \textit{doFlush} signal.
Given the fact that there is no mechanical shutter in the system, \textit{doFlush} is used to implement an electronic shutter. 
Indeed, the effective exposure time of the CCD, is the amount of time elapsed between the last flush request and the new acquisition request.

Once started, the FSM moves to the \textit{preVert} state, where it requests the removal of the optically covered lines to the `vertical timing' module, thanks to the \textit{startV} and \textit{shiftCount} signals.
As soon as the vertical shifting is done (\textit{shiftDone}), the inner state machine loop processes each horizontal line (states \textit{Vshift} to \textit{postHor}).
Likewise to the vertical shifting, optically covered pixels in the horizontal direction are removed before (\textit{preH}) and after (\textit{postH}) the optically effective zone (\textit{validH}).
Finally, when the full CCD processing is done, a \textit{CCD\_done} signal is generated to inform the `DAQ manager' module of the timing generation completion.

The PBLK signal, used to force the CCD processor data output at zero, is activated when the FSM is not moving on the inner loop, in other word, it is the complementary of HBLK.
CLOPB, used to remove residual offsets in the CCD processor signal chain,  is activated when reading the last optically covered pixels of each line (\textit{postH} pixels).
A \textit{valid\_pixel} signal is constructed to ease the tasks of the acquisition manager module (see section~\ref{daqManagerSec}).
This signal is activated only during the \textit{validHor} state and if the line currently accessed is in the allowed range determined by \textit{vertical start} and \textit{vertical stop}.
Indeed, the complete CCD sensitive zone can never be utilized for many reasons (alignment issues, exposed area, ...). 
The vertical selection was implemented to reduce the data volume by removing the meaningless lines.

The `ADC timing' modules, that are used to precisely control the CCD processor sampling times, are activated at start-up time by \textit{enable\_SHPD} and are free running.
Each module is composed of a FSM operated at 80\,MHz and a fast Dual Data Rate serializer (OSERDES) operated at 240\,MHz.
The FSM, which loops every eight 80\,MHz clock cycles (or steps), constructs and feeds the adequate 6 bit sub-step word to be serialized at 480\,Mbit/s at every steps.
Thus, a tuning resolution of about 2\,ns (or 10\,MHz/48) can be reached.
To set the start and stop times, 7 bit words are used.
The three MSB of the word select during which step the signal must be activated, while the four LSB select at which sub-step the activation will take place.

\begin{figure}[ht]
\begin{center}
\includegraphics[angle=0,width=0.9\textwidth]{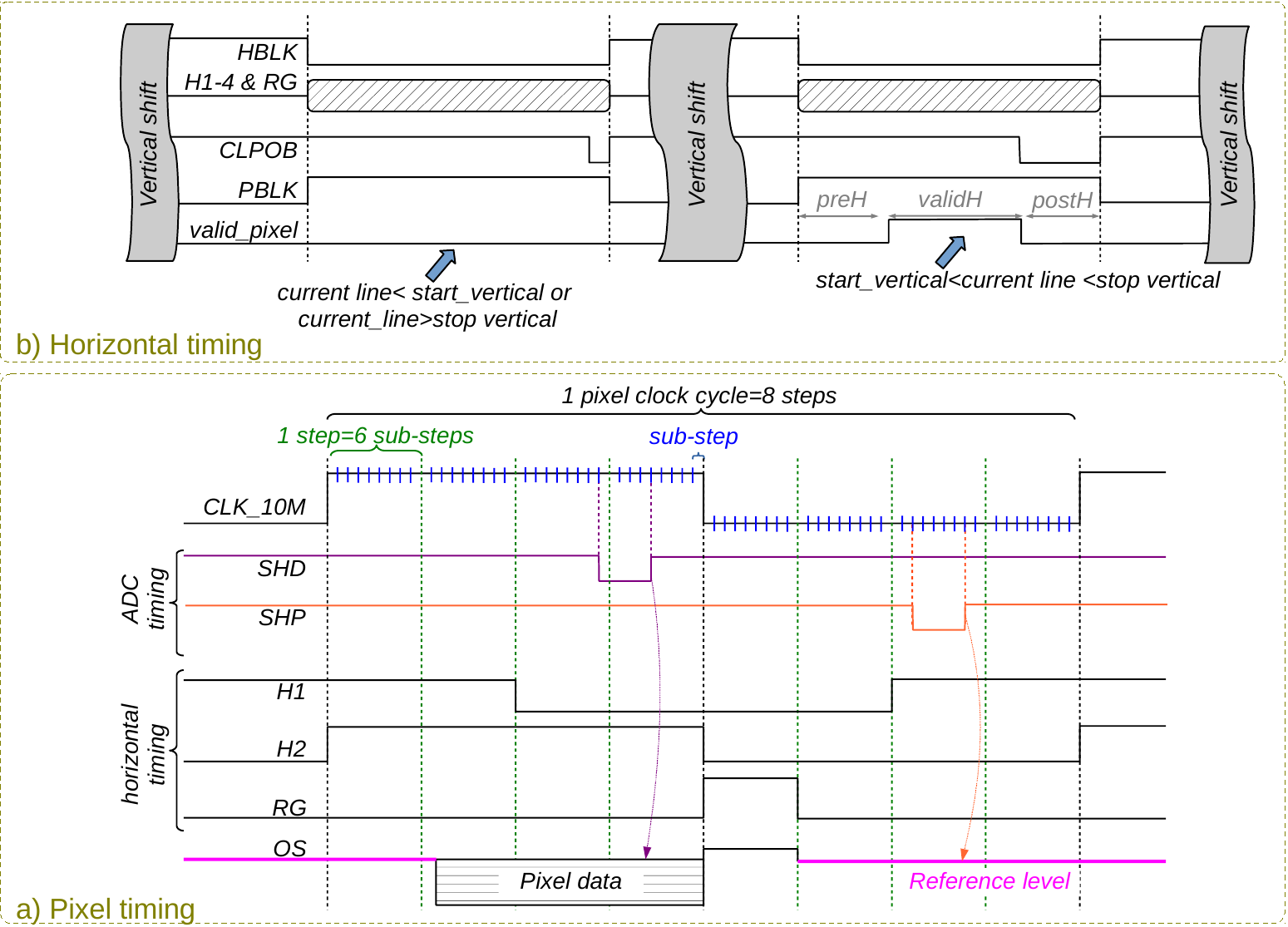}
\caption{Overview of the various signal timings involved in a CCD readout.
The upper part of the diagram, depicts the behavior of the signals controlled by the `timing core' inner FSM loop.
The lower part of the diagram, shows the details of the ADC timing and the horizontal timing within a pixel clock cycle.}
\label{timingDiag}
\end{center}
\end{figure}

Figure~\ref{timingDiag} gives an overview of the various signal timings involved in a CCD readout.
The upper part of the diagram, depicts the behavior of the signals controlled by the `timing core' inner FSM loop.
The lower part of the diagram, shows the details of the ADC timing and the horizontal timing within a pixel clock cycle.

\subsection{DAQ manager module description}
\label{daqManagerSec}
\begin{figure}[ht]
\begin{center}
\includegraphics[angle=0,width=0.9\textwidth]{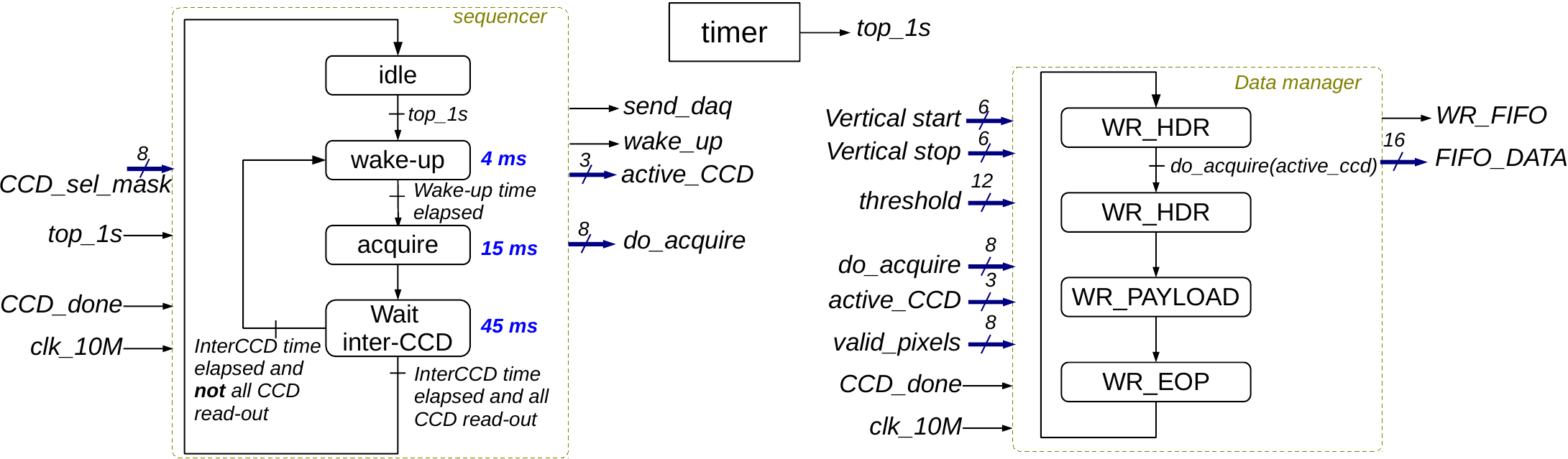}
\caption{Block diagram of the `DAQ manager' module.
The module is composed of two finite state machines (FSM): the acquisition sequencer (l.h.s) and the data manager (r.h.s).}
\label{daqDiag}
\end{center}
\end{figure}

A block diagram of the `DAQ manager' module is shown in figure~\ref{daqDiag}.
The module is composed of two finite state machines (FSM): the `acquisition sequencer' and the `data manager'.

The `acquisition sequencer' FSM is restarted every second by a timer.
Each CCD to be acquired  (selection through the selection mask signal), is read-out one after another.
At first, the `CCD processor' is waken-up via a request to the `serializer' module. This operation takes about 4\,ms.
Then the `CCD timing' module is started with the \textit{do\_acquire} signal (duration of about 15\,ms) and the FSM moves directly in the \textit{inter-CCD} waiting state.
The waiting timer is set to 60\,ms, but given the fact that a CCD readout takes 15\,ms, the real waiting time with no operation is indeed 45\,ms.
As soon as the CCD acquisition is done, a request is sent to the serializer to sleep the `CCD processor'.

The `data manager' FSM is started at each CCD acquisition request.
It first writes a data header, which contains useful information to determine if the data selection was used or not.
If data selection is used the 15 LSB data only contain the CCD number, otherwise, they additionally contain the \textit{vertical start} and \textit{vertical stop} parameters.
Indeed, when data selection is used, the valid pixel data words are not stored one after another but rather two data words per pixel, since their position in the payload do not reflect their coordinates anymore.
In that latter case, the two words data results from the concatenation of the coordinate word composed of 17 bit and of the pixel digitized value (12 bit).
Finally, once the full CCD payload is written in the memory, an End of Packet (EOP) is written in the FIFO memory.
The EOP is composed of one or two words, again depending on the acquisition mode.
In the full readout mode, the EOP word repeats the read-out CCD number, and the  `vertical start' and `vertical stop' parameters.
In the discrimination mode, the EOP two words contain the read-out CCD number and the number of pixels discriminated.
Naturally, the use of the data selection mode makes sense only for situations where less than half of the pixels contains data above threshold.

\section{Performances}
Several tests of the detector were performed, first with $\alpha$ particles of varying energies to check the energy response of the system, then with a cold neutron beam at the PF1B beam line at the Institute Laue Langevin (ILL). 

Particles are identified by looking at the pixel with the highest ADC value.
On the $11\times11$ pixels matrix centered on that pixel, three quantities are reconstructed~: the total sum of ADC values $\Sigma$ and the weighted averages $x$ and $y$ of the position in both directions.
This procedure is then repeated on the next remaining highest ADC value pixels until all particles have been reconstructed

\subsection{$\alpha$ measurement}
\begin{figure}[ht]
\begin{center}
\includegraphics[angle=0,width=0.49\textwidth]{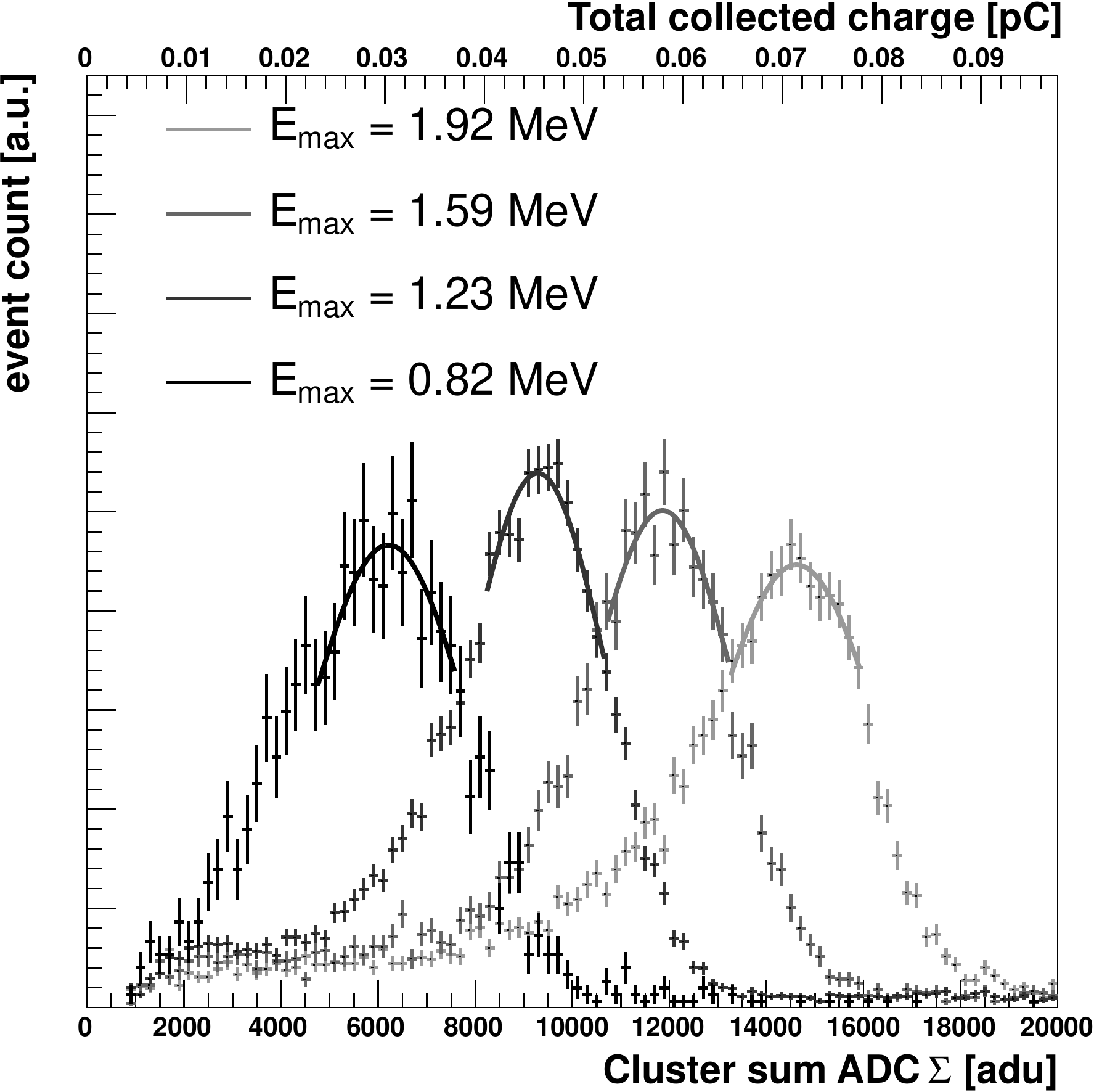}
\includegraphics[angle=0,width=0.49\textwidth]{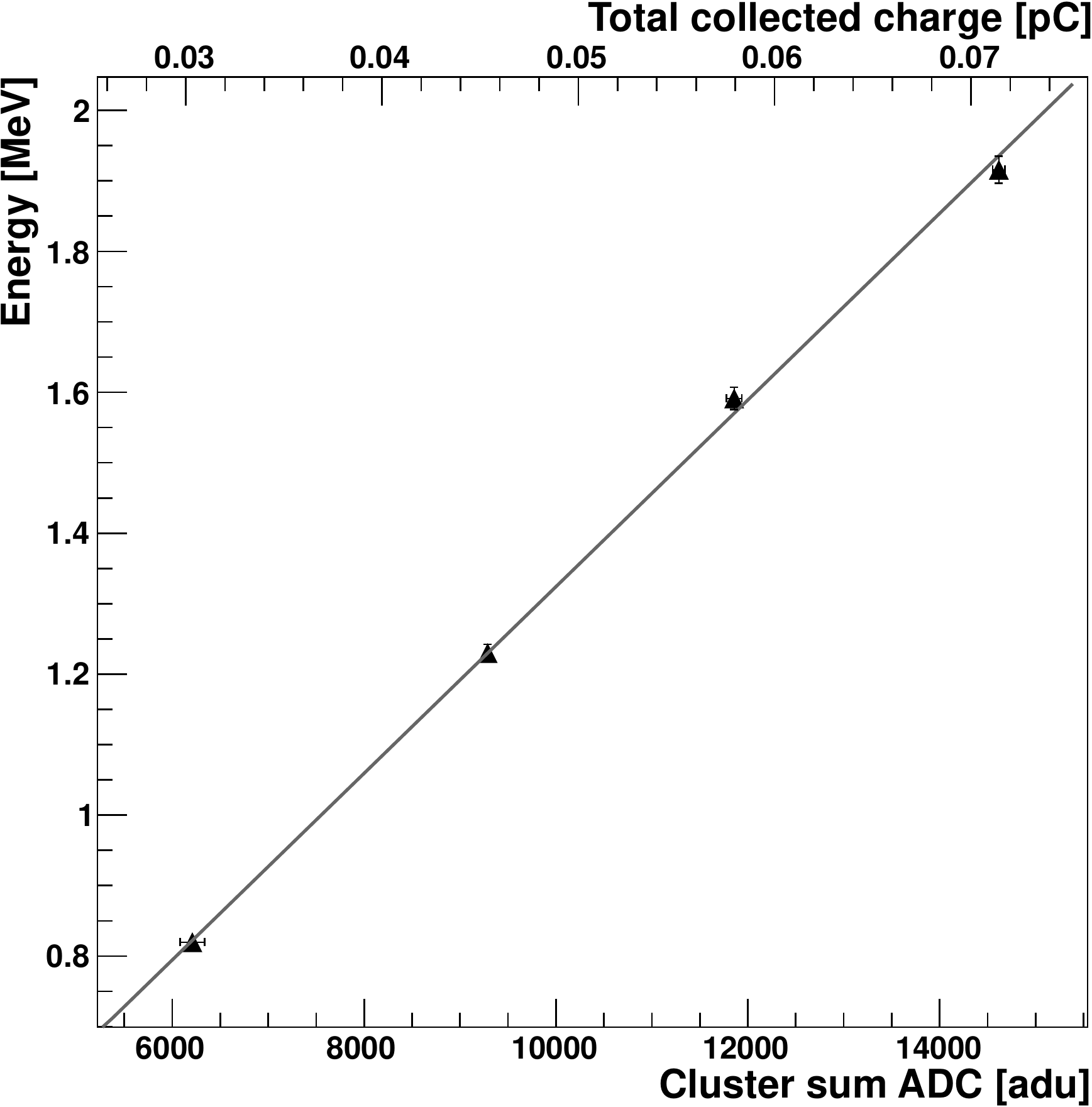}
\caption{Energy calibration and linearity: on the left the four energy spectra corresponding to different source-sensor distances and therefore to different peak energies; on the right the calibration curve obtained from the fit of the ADC peak.}
\label{fig-calib}
\end{center}
\end{figure}

Energy measurement is not in itself necessary for the use of the UCNBox detector. Nevertheless, measuring the energy resolution and the linearity of the sensor is a test of the efficiency of the system. 
To this aim, an \textsuperscript{241}Am $\alpha$ source was used.
The primary 5.48\,MeV particles were slowed down by a 12\,\textmu m aluminum foil.
The energy was further decreased by changing the distance between the source and the sensor in air at 1\,bar, thus allowing to reach energies between 1 and 2\,MeV.
This procedure widens significantly the energy distribution and reduces the precision of the measurement. 
For four different positions, the average energy was estimated using NIST tables~\cite{NISTtable}.
The resulting sum ADC spectra are presented in figure~\ref{fig-calib} with the resulting calibration curve.
The response is found to be linear, within the precision of this measurement.
The offset is set to zero, and the resulting fit gives a $\chi^2/Ndf = 2.13/3$, which is consistant with linearity.
The slope translates into a collected charge of $7550\pm50$\,adu/MeV$ = 0.0369\pm0.0002$\,pC/MeV, whereas one would expect a created charge of $0.0443$\,pC/MeV in pure silicon.
The difference can be accounted for by pair recombination within the CCD during the large exposition time (1\,s) and the clustering algorithm.
It does not impact the final performance as $\alpha$  particles from 0.8\,MeV to 2\,MeV are clearly identified.

\subsection{Neutron measurement}
\begin{figure}[ht]
\begin{center}
\includegraphics[angle=0,width=0.55\textwidth]{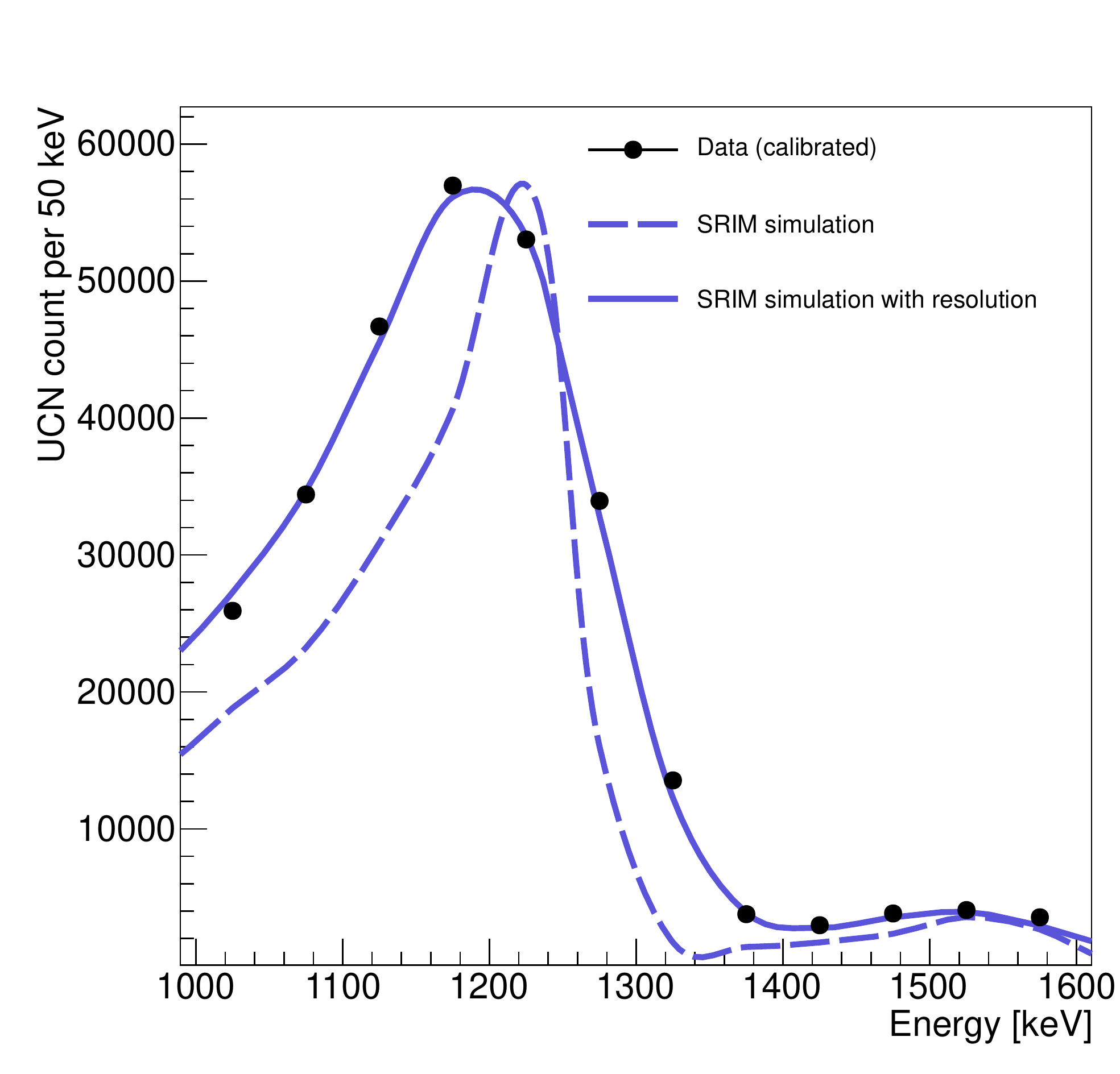}
\caption{Energy spectrum of $\alpha$ particles produced in the boron layer by neutron capture. The blue lines are the result of SRIM simulations of the setup with (plain line) or without (dashed line) accounting for detector resolution.}
\label{fig-neutron}
\end{center}
\end{figure}

The energy resolution was further investigated by exposing the detector to cold neutrons at the PF1B beam-line at ILL.
For this experiment, the detector was in a dark room, not fully isolated from ambient light. Therefore the pixel threshold was set relatively high. Simulation of the attenuation of $\alpha$ particles produced in the boron layer was performed using SRIM~\cite{SRIM}. The particle energy was determined using the previous calibration.
The comparison between reconstructed data and simulation, shown in figure~\ref{fig-neutron}, allows to extract an energy resolution of $58$\,keV. This value is sufficient for the purpose of the detector and will probably improve when lowering the pixel threshold.

\section{Summary}
\label{SummarySec}
To read-out a newly developed low energy neutron detector based on a set of 8 CCDs (sensitive area of  $ \rm 300 \, mm \times 0.8 \, mm$), a dedicated electronics was designed.
This electronics had to provide various features such as exposure time adjustment (0\,ms to 970\,ms), LED light calibration, embedded data-reduction, minimization of dead-time (\textless ~1\,\%) and low power usage (about 3\,W) to operate under vacuum. 
Additionally, the mechanical support had to provide a good thermal coupling to maintain the device at reasonable temperatures under vacuum and, at the same time, a precise adjustment mechanism to permit the relative height alignment of the CCDs. Using an $\alpha$ source of $^{241}$Am and a cold neutron beam, the performances of the full system (mechanical support, CCD and readout electronics) have been checked. In summary, this electronics is able to read-out simultaneously the eight CCDs at a rate of 1\,Hz and to meet all experimental requirements.


\end{document}